\documentclass[apl,superscriptaddress,
preprint,
]{revtex4-1}
\newcommand {\etal}{\textit{et al}.}
\usepackage{graphicx}
\usepackage{bm}
\usepackage{pifont}
\usepackage{amsmath}
\usepackage{amssymb}
\usepackage{easyReview}

\usepackage{hyperref}

\hypersetup{
 bookmarksnumbered=true,%
 bookmarksopen=true,%
 colorlinks=true,%
 linkcolor=blue,
 citecolor=blue,
 urlcolor=blue
}

\usepackage{color}
\usepackage{ulem}

\DeclareRobustCommand{\Erase}{\bgroup\markoverwith{\textcolor{red}{\rule[.5ex]{2pt}{0.4pt}}}\ULon}
\begin{document}
\title{
Novel supercell compounds of layered Bi-Rh-O with $p$-type metallic conduction materialized as a thin film form
}
%
\author{M. Ohno}
\affiliation{Department of Applied Physics and Quantum-Phase Electronics Center (QPEC), the University of Tokyo, Tokyo 113-8656, Japan}
\author{T. C. Fujita}
\email[Author to whom correspondence should be addressed: ]{fujita@ap.t.u-tokyo.ac.jp}
\affiliation{Department of Applied Physics and Quantum-Phase Electronics Center (QPEC), the University of Tokyo, Tokyo 113-8656, Japan}
\author{Y. Masutake}
\affiliation{Photon Factory, Institute of Materials Structure Science, High Energy Accelerator Research Organization (KEK), Tsukuba 305–0801, Japan}
\affiliation{Institute of Multidisciplinary Research for Advanced Materials (IMRAM), Tohoku University, Sendai 980-8577, Japan}
\author{H. Kumigashira}
\affiliation{Photon Factory, Institute of Materials Structure Science, High Energy Accelerator Research Organization (KEK), Tsukuba 305–0801, Japan}
\affiliation{Institute of Multidisciplinary Research for Advanced Materials (IMRAM), Tohoku University, Sendai 980-8577, Japan}
\author{M. Kawasaki}
\affiliation{Department of Applied Physics and Quantum-Phase Electronics Center (QPEC), the University of Tokyo, Tokyo 113-8656, Japan}
\affiliation{RIKEN Center for Emergent Matter Science (CEMS), Wako 351-0198, Japan}
%
\begin{abstract}
	Layered oxides have been intensively studied due to their high degree of freedom in designing various electromagnetic properties and functionalities.
	While Bi-based layered supercell (LSC) compounds [Bi$_n$O$_{n+\delta}$]-[$M$O$_2$] ($M$ = Mn, Mn/Al, Mn/Fe, or Mn/Ni; $n=2, 3$) are a group of prospective candidates, all of the reported compounds are insulators.
	Here, we report on the synthesis of two novel metallic LSC compounds [Bi$_{n}$O$_{n+\delta}$]-[RhO$_2$] ($n=2, 3$) by pulsed laser deposition and subsequent annealing.
	With tuning the thickness of the sublattice from Bi$_2$O$_{2+\delta}$ to  Bi$_3$O$_{3+\delta}$, a dimensionality-dependent electrical transport is revealed from a conventional metallic transport in $n=2$ to a localized transport in $n=3$.
	Our successful growth will be an important step for further exploring novel layered oxide compounds.
\end{abstract}
\maketitle

\begin{figure}
	\begin{center}
		\includegraphics*[bb=0 0 304 276,width=13cm]{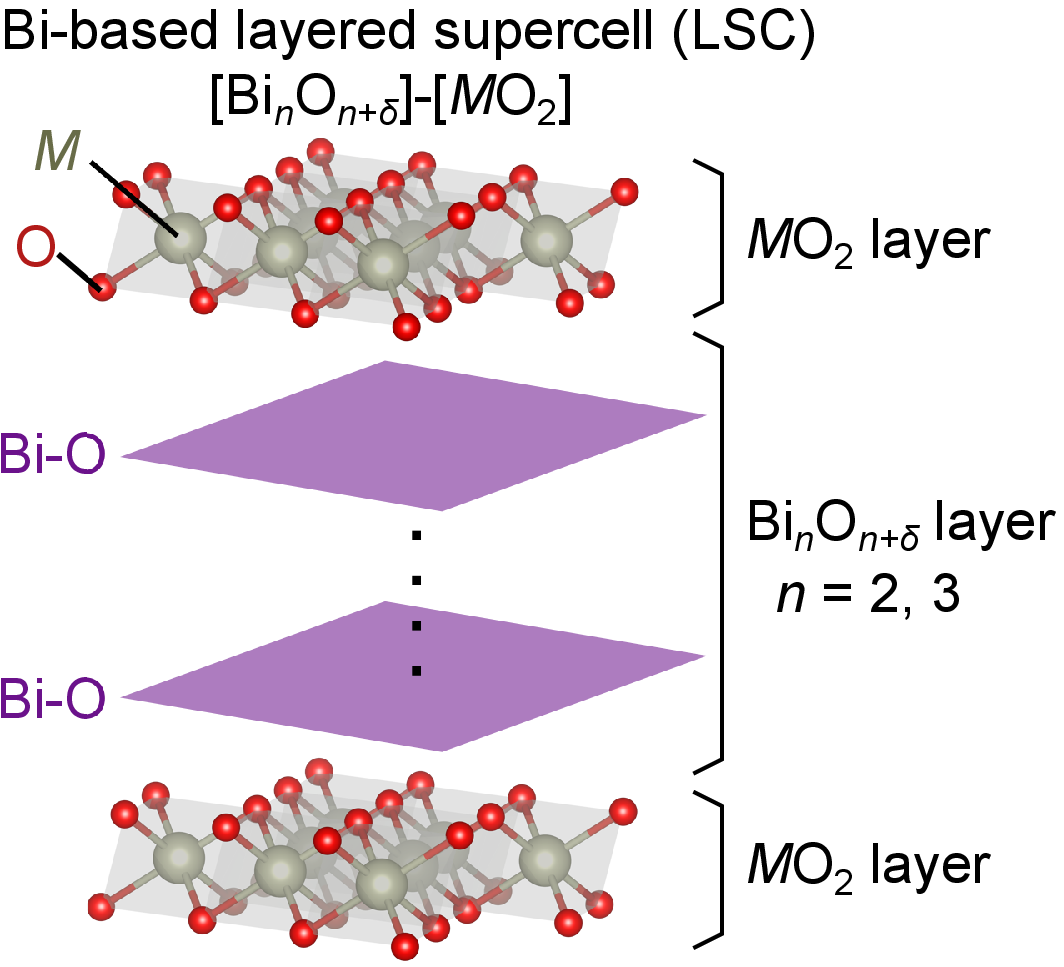}
		\caption{
			Schematic of Bi-based layered supercell (LSC) structure, namely [Bi$_n$O$_{n+\delta}$]-[$M$O$_2$].
		}
		\label{fig1}
	\end{center}
\end{figure}

Layered compounds have attracted growing attention over the past decades due to their tunability of physical properties and enormous potential in nanoscale devices~\cite{Geim2007, Bonaccorso2010, Schwierz2010, Novoselov2012,Wang2012h,Yu2013,Jena2014,Wang2015c,Wen2016}.
Above all, layered oxides open up great flexibility and opportunities in structure construction as well as multifunctionality exploration due to highly tunable crystal structures and controllable degrees of freedom~\cite{Osada2012,Raveau2013,Butler2013,Kalantar-zadeh2016}.
With the development of thin film growth techniques, various layered oxides with designed crystal structures have been fabricated~\cite{Schaak2002,Battle1997,Birenbaum2014,Chen2013a}.
Recently, a series of Bi-based layered supercell (LSC) structures with a formula [Bi$_n$O$_{n+\delta}$]-[$M$O$_2$] ($M$ = Mn, Mn/Al, Mn/Fe, and Mn/Ni; $n=2, 3$) has been discovered only in a thin film form synthesized by a pulsed laser deposition (PLD) technique~\cite{Li2017c,Li2019o,Jin2020a,Misra2020,Shen2022}.
While it is generally difficult to identify such misfit structures in a thin-film form, alternative layer stacking of two sublattices of Bi$_n$O$_{n+\delta}$ and $M$O$_2$ along the film growth direction is confirmed in all the examples, as shown in Fig.~\ref{fig1}.
Because of magnetic $3d$ transition elements, Bi-based LSC structures exhibit high ferromagnetic Curie temperatures above 300~K and highly anisotropic magnetic and optical properties, which endow them with huge potential in various application fields.
Furthermore, by precisely controlling the growth conditions, various LSC structures can be synthesized by changing the thickness of the Bi-O layers~\cite{Jin2020a,Li2019o}.
Due to this structural flexibility, LSC compounds exhibit tunable physical properties, for example, [Bi$_3$O$_{3+\delta}$]-[MnO$_2$] have three times higher out-of-plane magnetization and 10\% higher optical transmittance than [Bi$_2$O$_{2+\delta}$]-[MnO$_2$]~\cite{Jin2020a}.

So far, all of the reported Bi-based LSC compounds are insulators.
In this report, in order to apply their anisotropic and tunable physical flexibility to electrical functionalities, we replaced previously studied $3d$ elements with Rh, which has $4d$ electrons with a more spatially extended nature.
Here, we report the synthesis of two new LSC compounds [Bi$_{n}$O$_{n+\delta}$]-[RhO$_2$] ($n = 2, 3$) (named as 2-Bi and 3-Bi hereafter).
As expected, both 2-Bi and 3-Bi exhibit a $p$-type and metallic electrical transport that has not been reported so far for LSC compounds.
Furthermore, they show a peculiar Bi-O layer-number-dependency plausibly related to the spatial dimensionality of the conducting RhO$_2$ layer.
Detailed x-ray photoemission spectroscopy (XPS) reveals that the top of the Rh $4d$ valence band dominantly contributes to the $p$-type metallic electrical transport due to hole doping into Rh$^{3+}$-based semiconducting states.

2-Bi and 3-Bi films are synthesized on Y-stabilized ZrO$_2$ (YSZ) (111) substrates by the deposition of amorphous films at room temperature by PLD and subsequent annealing.
Before the film growth, YSZ (111)  substrates were annealed in air with a muffle furnace at 1,350~${^\circ}$C for 3~h to obtain a clear step-terrace structure with single-unit-cell height ($\sim$3 {\AA}).
A polycrystalline PLD target of Bi$_2$Rh$_2$O$_7$ was prepared by mixing Bi$_2$O$_3$ and Rh$_2$O$_3$ powders at a molar ratio of Bi:Rh =1:1, pelletizing and sintering it at 950~${^\circ}$C for 24~h in a tube furnace under O$_2$ flow (200~mL/min.)~\cite{Longo1972,Kennedy1997,Li2013}.
The films were deposited at room temperature under $5\times10^{-7}$~Torr O$_2$ by KrF excimer laser ($\lambda=248$~nm) pulses at a frequency of 5~Hz and an energy fluence of 1.1~J/cm$^2$.
The deposition rate was about 0.2~$\mathrm{\AA}$/s.
3-Bi and 2-Bi films were synthesized respectively by annealing at 700 and 900~${^\circ}$C for 1~h in a tube furnace under O$_2$ flow (200~mL/min.).
Structural properties of the films were characterized by x-ray diffraction (XRD) (SmartLab, Rigaku) and cross-sectional transmission electron microscopy at room temperature.
The magnetotransport properties were measured with using a cryostat equipped with a 9~T superconducting magnet (PPMS, Quantum Design Co.).
XPS measurements were performed using a Scienta Omicron R3000 analyzer with a monochromatized Al K$_\alpha$ x-ray source ($h\nu=1486.6$~eV).
Binding energies were calibrated by measuring a gold film electrically connected to the samples.
All spectra were acquired at room temperature with a total energy resolution of 500~meV.

\begin{figure}
	\begin{center}
		\includegraphics*[bb=0 0 431 383,width=15cm]{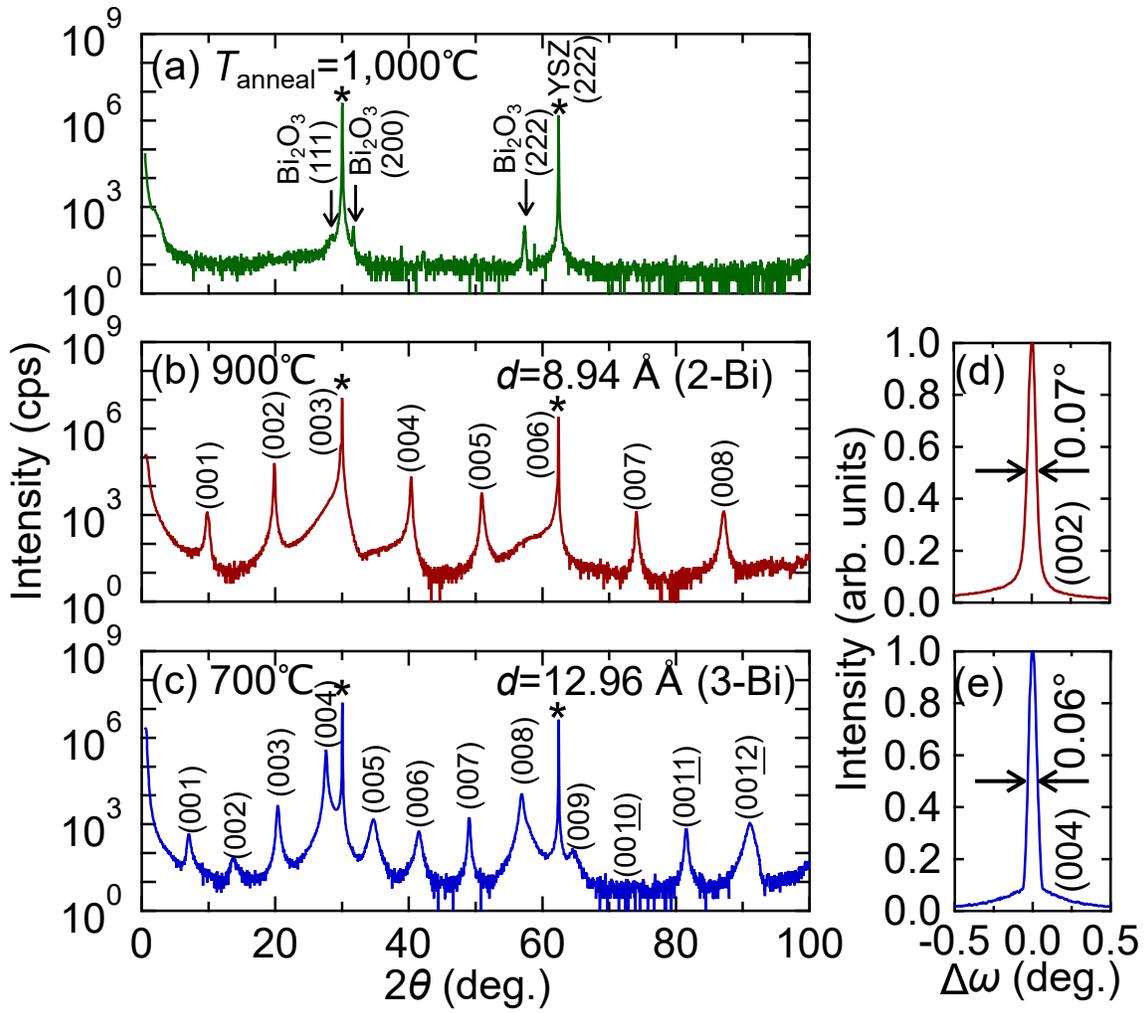}
		\caption{
			X-ray diffraction (XRD) patterns for Bi-Rh oxide films annealed at various temperatures ($T_\mathrm{anneal}$). 
			XRD $\theta$-2$\theta$ scans after annealing at (a) 1,000, (b) 900, and (c) 700~${^\circ}$C.
			YSZ substrate peaks are marked with asterisks.
			Rocking curves are shown around the (d) (002) peak in (b), and (e) (004) peak in (c).
		}
		\label{fig2}
	\end{center}
\end{figure}

Figs.~\ref{fig2}(a)--(c) summarize an annealing temperature ($T_\mathrm{anneal}$) dependence of XRD $\theta$-$2\theta$ scan.
As shown in Fig.~\ref{fig2}(a), Bi$_2$O$_3$ is formed at high temperature ($T_\mathrm{anneal}=1,000$~${^\circ}$C).
In contrast, as shown in Figs.~\ref{fig2}(b) and \ref{fig2}(c), the peaks are assignable to the phase-pure compounds with (001) growth orientation.
The two kinds of periodic peaks correspond to the $d$-spacings of $8.94$~{\AA} ($T_\mathrm{anneal}=900$~${^\circ}$C) and $12.96$~{\AA} ($T_\mathrm{anneal}=700$~${^\circ}$C), both of which are not found in the ICDD powder diffraction database of the Bi-Rh-O system.
Such large $d$-spacings are similar to those of other LSC compounds~\cite{Li2017c,Li2019o,Jin2020a,Misra2020,Shen2022}.
Also similarly, the two large $d$-spacings with an obvious difference suggest that the LSC structures are formed with two different stacking sequences depending on the growth conditions~\cite{Jin2020a,Li2019o}.
Rocking curves taken for the (002) film peak in Fig.~\ref{fig2}(d) and the (004) film peak in Fig.~\ref{fig2}(e) are very sharp with full widths at half maximum of $0.07$~degrees and $0.06$~degrees, respectively, ensuring high crystallinity of the films.

\begin{figure*}
	\begin{center}
		\includegraphics*[bb=0 0 2136 781,width=17cm]{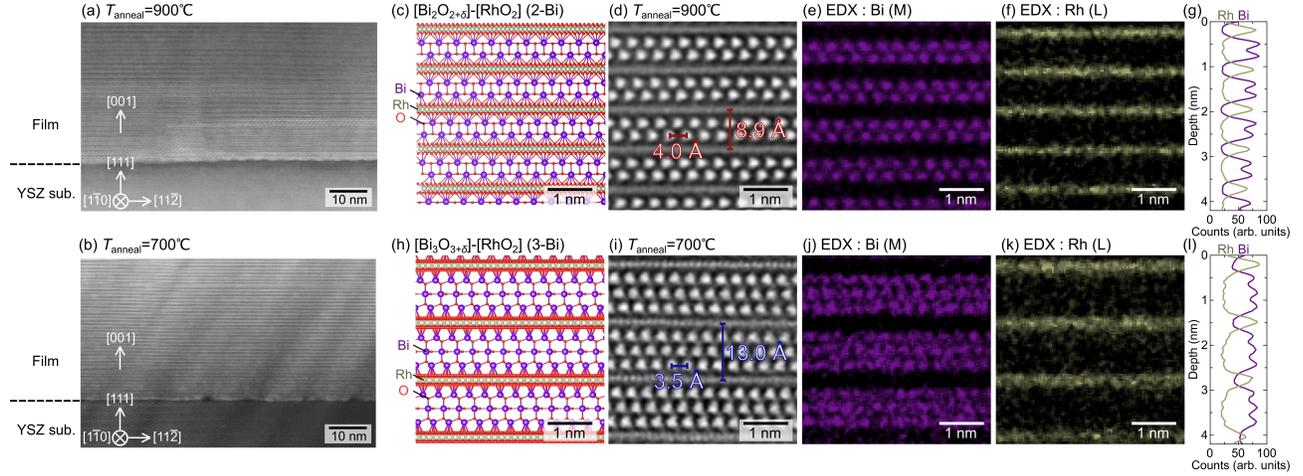}
		\caption{
			Wide-range cross-sectional high-angle annular dark-field (HAADF) scanning transmission electron microscopy (STEM) image of the films annealed at (a) 900 and (b) 700~${^\circ}$C.
			Schematics of atomic arrangement for (c)  [Bi$_2$O$_2$]-[RhO$_2$] (2-Bi) and (h) [Bi$_3$O$_3$]-[RhO$_2$] (3-Bi) stacking structures.
			Higher resolution HAADF-STEM images of the films annealed at (d) 900~${^\circ}$C and (i) 700~${^\circ}$C viewed along the direction where Bi atoms can be clearly resolved.
			The corresponding energy dispersive x-ray (EDX) spectrometry maps for (e, j) Bi M and (f, k) Rh L edges in (d) and (i), respectively.
			Depth profile of the signal for Bi and Rh obtained by integrating the EDX counts along the horizontal direction of (e), (f) in (g) and (j), (k) in (l).
		}
		\label{fig3}
	\end{center}
\end{figure*}

In order to clarify the microscopic structures of the new layered compounds revealed by the XRD, we performed cross-sectional scanning transmission electron microscopy (STEM) measurements as shown in Fig.~\ref{fig3}.
It is clarified that the new compounds indeed have LSC structures schematically shown in Figs.~\ref{fig3}(c) and \ref{fig3}(h).
Wide-range high-angle annular dark-field (HAADF)-STEM images taken for the films annealed at 900 (Fig.~\ref{fig3}(a)) and 700~${^\circ}$C (Fig.~\ref{fig3}(b)) show clear contrast, confirming the formation of layered structures.
We found that the in-plane axes of the films are oriented rather randomly, as confirmed by the presence of multiple domain structures.
Dotted and continuous lines in bright layers are indicative of channeling and non-channeling conditions of incident electron beam, respectively (see the supplementary material).
Figure~\ref{fig3}(d) presents a higher resolution cross-sectional HAADF-STEM image of 900~${^\circ}$C annealed film taken by tuning the rotating stage to observe the channeling of Bi atoms.
The STEM image shows a periodic stack of two dotted lines and one continuous line with a period of 8.9~{\AA}, which is in good agreement with the out-of-plane $d$-spacing value determined by XRD.
Energy dispersive x-ray spectrometry (EDX) maps for the same area are given in Figs.~\ref{fig3}(e)--~\ref{fig3}(g), clarifying that the dotted lines and the continuous line correspond to the Bi bilayer and the Rh monolayer, respectively.
The above result indicates that 900~${^\circ}$C annealed film possesses 2-Bi structure, with an alternative layered stacking of sublattices composed of Bi$_2$O$_{2+\delta}$ and RhO$_2$ along the film growth direction as shown in Fig.~\ref{fig3}(c)~\cite{Li2017c,Li2019o}.
The same dataset for 700~${^\circ}$C annealed film is presented in Figs.~\ref{fig3}(i)--~\ref{fig3}(l). 
They show a similar periodic stack of Bi and Rh layers except for the additional insertion of Bi-O layer.
As a result, the periodicity along the growth direction is 13.0~{\AA} that is again in good agreement with the out-of-plane $d$-spacing value determined by XRD.
These results indicate that 700~${^\circ}$C annealed film corresponds to 3-Bi structure as shown in Fig.~\ref{fig3}(h).
From the above microstructural characterization and analysis, it is concluded that the layered stacking sequences of Bi-based layered structures can be tuned by controlling the annealing temperature, similar to other LSC compounds~\cite{Li2019o,Jin2020a}.
Here, it is worth mentioning that the determination of the oxygen atom position and composition generally requires intricate measurements and analyses such as precession electron diffraction tomography that is employed for other LSC structures~\cite{Li2017c,Li2019o}.
Although it is intriguing to elucidate the precise crystal structures of the new compounds, this is beyond the scope of this report and remains as future work.
However, because of the similarity of the structures observed in the STEM measurements with those of previously reported other LSC, we surmise that  [Bi$_n$O$_{n+\delta}$]-[RhO$_2$] ($n = 2, 3$) structures are quite possibly materialized.
A further analysis of misfit layered structures, focusing on the distinct stacking configurations of the Bi-O layers in 2-Bi and 3-Bi films, is provided in the supplementary material.

Figure~\ref{fig4} summarizes fundamental electrical transport properties of 2-Bi and 3-Bi films.
The temperature dependence of the longitudinal resistivity $\rho_{\mathrm{xx}}$ is presented in Fig.~\ref{fig4}(a), where no phase transitions are observed between 2 and 300~K.
While 2-Bi film exhibits a metallic behavior down to 2~K with a residual resistivity ratio [RRR = $\rho_{\mathrm{xx}}$(300~K)/$\rho_{\mathrm{xx}}$(2~K)] of 6.0, 3-Bi film shows a minimum at about 40~K and a slight upturn upon further cooling (RRR $=1.6$), as shown in Fig.~\ref{fig4}(b).
Hence, a conventional metallic transport is observed in 2-Bi film and a localized transport in 3-Bi film, which may reflect the structural differences.
Figure~\ref{fig4}(c) presents magnetoresistance (MR) ratio taken with out-of-plane magnetic field swept at 2~K.
While 2-Bi film exhibits a quadratic MR, 3-Bi film shows a rather linear MR, indicating that localization occurred in 3-Bi film due to enhanced 2D nature. These behavior suggest that the RhO$_2$ and Bi$_{n}$O$_{n+\delta}$ layers can be viewed as conducting and blocking layers, respectively.
Fig.~\ref{fig4}(d) shows the Hall resistivity measured at 2~K.
The positive linear Hall resistivity is observed for both 2-Bi and 3-Bi films, indicating  a $p$-type carrier conduction.
The carrier density and mobility of 2-Bi (3-Bi) are estimated as 4 (6)$\times 10^{21}$~cm$^{-3}$ and 4 (1)~cm$^2$/Vs by single-carrier fitting.
Contrary to 2-Bi and 3-Bi films, another example of layered Bi-Rh oxide, Bi$_8$Rh$_7$O$_{22}$ film shows a semiconducting behavior (Fig.~\ref{fig4}(a)) and lower hole carriers ($2\times 10^{21}$~cm$^{-3}$ at 300~K)~\cite{Uchida2016} than 2-Bi and 3-Bi films, as listed in Table~I.
$Ex$-$situ$ annealing under strong oxidization conditions may introduce holes in filled Rh$^{3+}$ ([Kr] $4d^6$) states, which results in an increase of \textit{p}-type carriers and induce the metallic behavior of 2-Bi and 3-Bi film.

In order to understand the metallic and $p$-type transport properties, we examine the electronic structure of 2-Bi and 3-Bi films by using XPS, as shown in Fig.~\ref{fig5}.
Figure~\ref{fig5}(a) shows the Rh $3d$ core-level spectra of 2-Bi and 3-Bi films in comparison with those of Rh$_2$O$_3$ and RhO$_2$, which are references to Rh$^{3+}$ and Rh$^{4+}$ oxidation states, respectively~\cite{Abe2001}.
The binding energies of peaks in the Rh $3d$ spectra in 2-Bi and 3-Bi films are in excellent agreement with that of Rh$_2$O$_3$ (Rh$^{3+}$).
This indicates that 2-Bi and 3-Bi films predominantly consist of Rh$^{3+}$ oxidation states, where Rh $4d$ filled $t_{2g}$ and empty $e_\mathrm{g}$ manifolds serve as the valence and conduction bands, respectively. 
The almost filled Rh $4d$ $t_{2\mathrm{g}}$ states are confirmed by the valence-band spectra displayed in Fig.~\ref{fig5}(b); the prominent features in the energy range from $3$~eV to $E_\mathrm{F}$ mainly originated from Rh $4d$ states, while the broad features in the energy range from $8$ to $3$ eV mainly come from Bi $6p$ states.
A more detailed inspection of the spectra near $E_\mathrm{F}$ reveals the existence of small but finite DOS near $E_\mathrm{F}$.
This indicates the occurrence of slight hole-doping in the Rh $4d$ valence band and the dominant contribution of Rh $4d$ states to the $p$-type metallic conduction in 2-Bi and 3-Bi films.

In summary, we have fabricated two new LSC films with different numbers of Bi-O layers by controlling $T_\mathrm{anneal}$, starting from the same amorphous films prepared by PLD.
The combination of XRD and STEM has confirmed the formation of LSC structures with a layered stacking of [Bi$_n$O$_{n+\delta}$] ($n=2, 3$) and [RhO$_2$] sublattices along the film growth direction.
By substituting Rh to $M$O$_2$ layer, $p$-type metallicity is successfully introduced into LSC systems.
While 2-Bi film shows a conventional metallic transport, 3-Bi film a weakly localized transport, suggesting that the dimensionality can be tuned by changing the number of Bi-O layers.
From XPS measurements, the top of the Rh $4d$ valence band dominantly contributes to the $p$-type metallic electrical transport as a result of hole doping into Rh$^{3+}$ states.
Our successful growth of two conducting LSC films paves the way for further research into the functionality and structural flexibility of layered oxides. 

\sloppy
This work was supported by JSPS Grants-in-Aid for Scientific Research (S) No. JP22H04958, by JSPS Grant-in-Aid for Early-Career Scientists No. JP20K15168, by JSPS Fellowship No. JP22J12905, and by The Murata Science Foundation, Mizuho Foundation for the Promotion of Sciences, Iketani Science and Technology Foundation, The Kazuchika Okura Memorial Foundation, and Mitsubishi Foundation.
%


\clearpage


\begin{figure}
	\begin{center}
		\includegraphics*[bb=0 0 404 379,width=16cm]{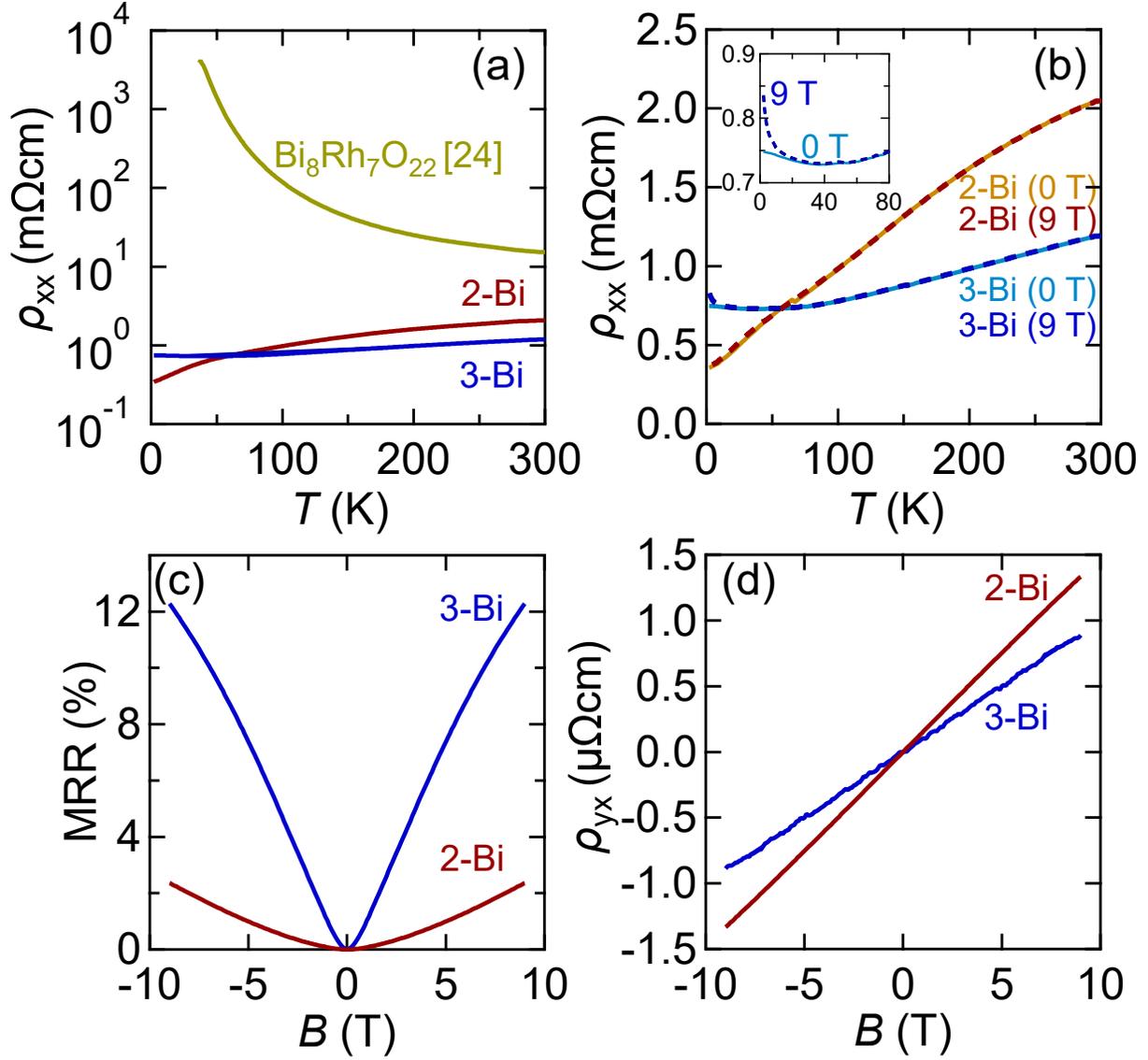}
		\caption{
			(a) Temperature dependence of longitudinal resistivity $\rho_{\mathrm{xx}}$ in 2-Bi and 3-Bi films, compared with that for previously reported Bi$_8$Rh$_7$O$_{22}$ film \cite{Uchida2016}.
			Magnetotransport properties of 2-Bi and 3-Bi films in (b)--(d).
			(b) Temperature dependence of $\rho_{\mathrm{xx}}$.
			The inset shows an enlarged view at low temperatures.
			(c) Magnetoresistance ratios (MRR $=(R_{\mathrm{xx}}(B)-R_{\mathrm{xx}}(0$~$\mathrm{T}))/R_{\mathrm{xx}}(0$~$\mathrm{T})$)
			taken with sweeping the out-of-plane field at 2~K.
			(d) Hall resistivity $\rho_{\mathrm{yx}}$ taken at 2~K.
		}
		\label{fig4}
	\end{center}
\end{figure}

\begin{figure}
	\begin{center}
		\includegraphics*[bb=0 0 501 380,width=15cm]{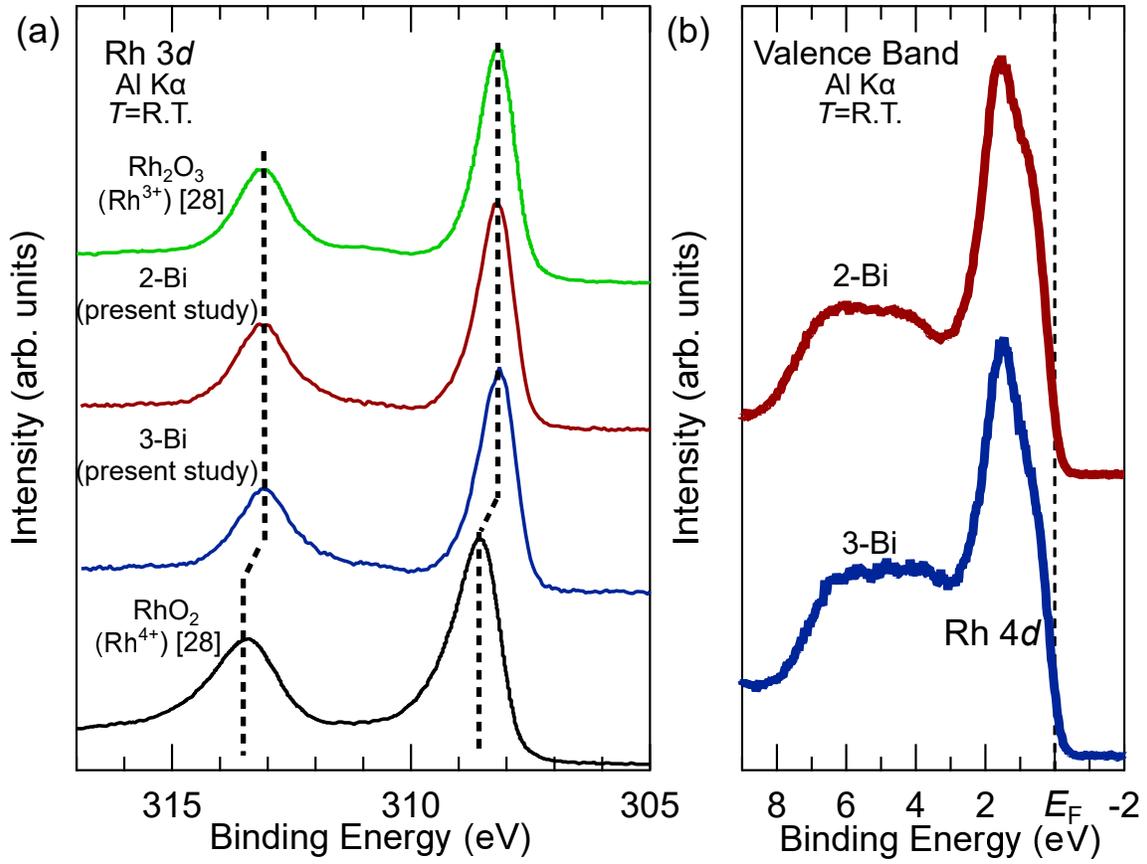}
		\caption{
			(a) Rh $3d$ core-level spectra for 2-Bi and 3-Bi films in comparison with Rh$_2$O$_3$ (Rh$^{3+}$) and RhO$_2$ (Rh$^{4+}$) bulks~\cite{Abe2001} as references.
			(b) Valence-band spectra of 2-Bi (red line) and 3-Bi (blue line) films.
		}
		\label{fig5}
	\end{center}
\end{figure}

\clearpage
\clearpage
\begin{table}[htbp]
	\caption{Growth conditions and transport properties of Bi-Rh oxide films.}
	\begin{ruledtabular}
		\begin{tabular}{ccccc}
			Compounds&Growth method& $\frac{\rho_{\mathrm{xx}}(300~\mathrm{K})}{\rho_{\mathrm{xx}} (2~\mathrm{K})}$&\multicolumn{2}{c}{$p$ ($10^{21}$~cm$^{-3}$)} \\ 
			&&& [300~K] & [2~K] \\ \hline 
			Bi$_8$Rh$_7$O$_{22}$~\cite{Uchida2016} & as deposition  & $<10^{-3}$ & 2 & --- \\ \relax
			&at 800~${^\circ}$C& & & \\ \relax
			[Bi$_2$O$_{2+\delta}$]-[RhO$_2$] &  $ex$-$situ$ annealing& 6.0& 4&4  \\ \relax
			(2-Bi)&at 900~${^\circ}$C&&& \\ \relax
			[Bi$_3$O$_{3+\delta}$]-[RhO$_2$] & $ex$-$situ$ annealing & 1.6&8&6 \\ \relax
			(3-Bi)&at 700~${^\circ}$C&&&
		\end{tabular}
	\end{ruledtabular}
\end{table}

\end{document}


\title{
		Supplementary Materials	 for\\
		 Novel supercell compounds of layered Bi-Rh-O with $p$-type metallic conduction materialized as a thin film form
	}
	%
	\author{M. Ohno}
	\affiliation{Department of Applied Physics and Quantum-Phase Electronics Center (QPEC), the University of Tokyo, Tokyo 113-8656, Japan}
	\author{T. C. Fujita}
	\email[Author to whom correspondence should be addressed: ]{fujita@ap.t.u-tokyo.ac.jp}
	\affiliation{Department of Applied Physics and Quantum-Phase Electronics Center (QPEC), the University of Tokyo, Tokyo 113-8656, Japan}
	\author{Y. Masutake}
	\affiliation{Photon Factory, Institute of Materials Structure Science, High Energy Accelerator Research Organization (KEK), Tsukuba 305–0801, Japan}
	\affiliation{Institute of Multidisciplinary Research for Advanced Materials (IMRAM), Tohoku University, Sendai 980-8577, Japan}
	\author{H. Kumigashira}
	\affiliation{Photon Factory, Institute of Materials Structure Science, High Energy Accelerator Research Organization (KEK), Tsukuba 305–0801, Japan}
	\affiliation{Institute of Multidisciplinary Research for Advanced Materials (IMRAM), Tohoku University, Sendai 980-8577, Japan}
	\author{M. Kawasaki}
	\affiliation{Department of Applied Physics and Quantum-Phase Electronics Center (QPEC), the University of Tokyo, Tokyo 113-8656, Japan}
	\affiliation{RIKEN Center for Emergent Matter Science (CEMS), Wako 351-0198, Japan}

	\maketitle
	\thispagestyle{plain}
\pagestyle{plain}
	
	%
	%
	\section{Lattice networks of Bi-O layers}
	Magnified HAADF-STEM images of 2-Bi and 3-Bi films are shown in Figs.~S1(c) and S1(f), respectively.
	They are similar to those of reported Bi-based LSC films [Bi$_n$O$_{n+\delta}$]-[$M$O$_2$] ($M$ = Mn, Mn/Al, Mn/Fe or Mn/Ni; $n=2, 3$)~\cite{Li2017c, Li2019o, Jin2020a, Misra2020, Shen2022}.
	As mentioned in the main text, the stacking of Bi-O layers in our 2-Bi and 3-Bi films are different from each other,  which has been commonly observed in previously reported Bi-based LSC films as well.
	Here, we propose that this difference may reflect distinct lattice arrangements of Bi-O layer; square (Fig.~S1(a)) for $n=2$  and triangular (Fig.~S1(d)) for $n=3$, which can reasonably reproduce the observed channeling of Bi atoms in STEM measurement.

	We focus on the incident electron beam direction where the channeling of Bi atoms are seen. 
	In the case of a square lattice (Fig.~S1(a)), Bi atoms in upper layer are located at the center of underlying Bi atoms (Figs.~S1(b), S1(c)), while in a triangular lattice (Fig.~S1(d)), Bi atoms are displaced from the center (Figs.~S1(e), S1(f)).
	As shown in Fig.~S1(c), Bi atoms in upper layer are located at the midpoint of underlying Bi atoms in 2-Bi film, akin to the square lattice arrangements observed in misfit oxides with CoO$_2$ layers~\cite{Nagai2006}.
	On the other hand, as shown in Fig.~S1(f), Bi atoms in upper layer are shifted from the midpoint of underlying Bi atoms in 3-Bi film, similar to the estimated structures of other Bi-based LSC compounds with $n=3$~\cite{Li2017c, Li2019o}.
	The presence of in-plane random domains in our films makes it challenging to determine their crystal structures.
	Additional studies are needed to estimate crystal structure of our films.

	\begin{figure}[h]
		\begin{center}
			\includegraphics*[bb=0 0 648 404,width=14cm]{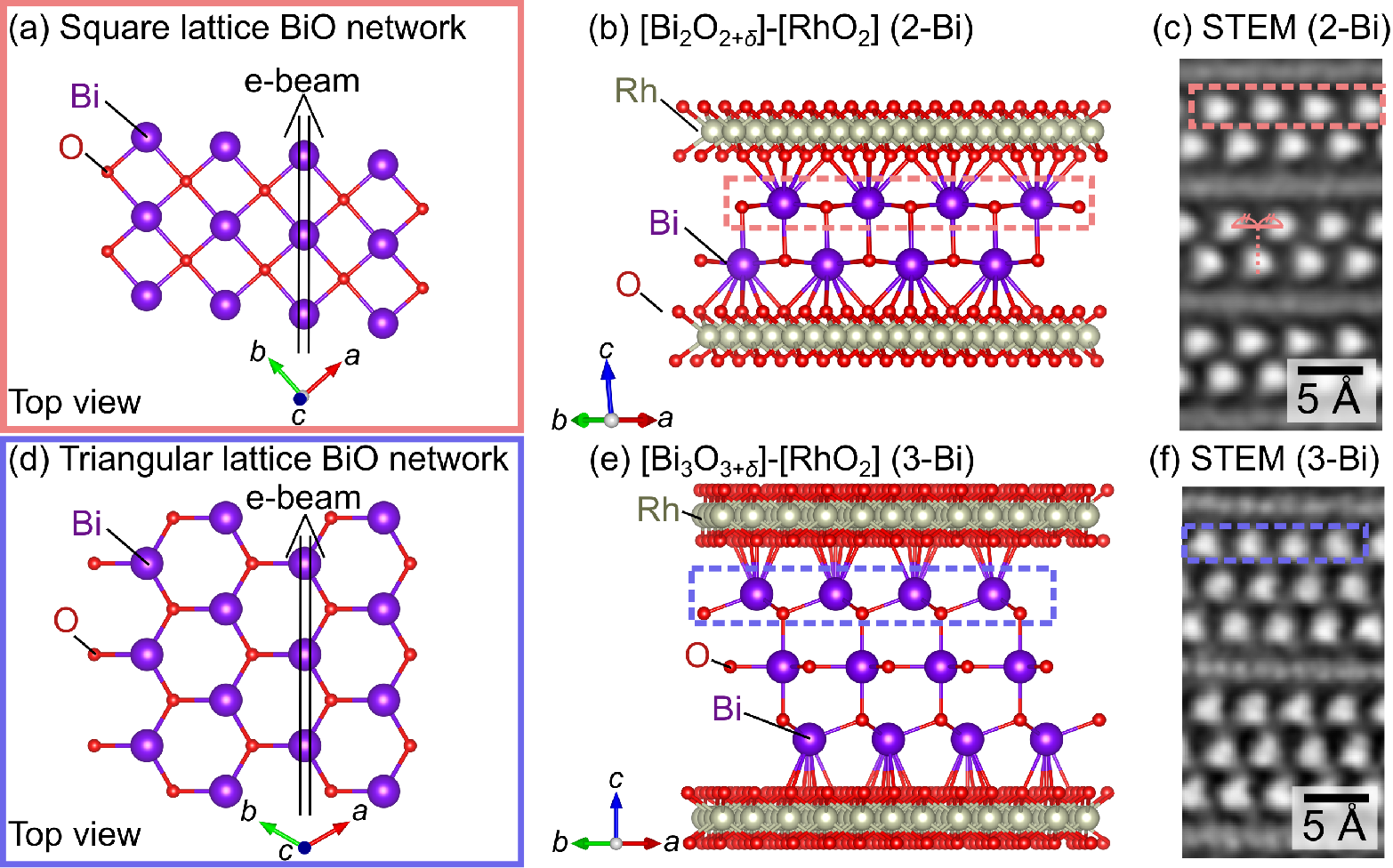}
			\caption{
				Schematics of top view Bi-O layers with (a) square and (d)  triangular lattice networks together with the direction of the e-beam for STEM measurements.
				Cross sectional schematics of (b) [Bi$_{2+\delta}$O$_2$]-[RhO$_2$] (2-Bi)and (e)  [Bi$_{3+\delta}$O$_3$]-[RhO$_2$] (3-Bi), viewed along the e-beam direction in (a) and (d).
				Red in (b) and blue in (e) dashed box areas of Bi-O layers correspond to triangular and square lattice network in (a) and (d), respectively.
				Schematics for 2-Bi in (a) and (b) are sketched with referring the structure of [Sr$_2$O$_2$]$_{0.5}$CoO$_2$~\cite{Nagai2006} with square lattice network of SrO layers.
				Schematics for 3-Bi in (d) and (e) are reproduced from the structure of  [Bi$_3$O$_{3+\delta}$]-[MnO$_2$]~\cite{Li2017c, Li2019o}.
				Higher resolution high-angle annular dark-field scanning transmission electron microscopy images of  (c)  2-Bi and (f) 3-Bi films.
			}
			\label{figS1}
		\end{center}
	\end{figure}
